\documentstyle[aps,multicol,eqsecnum,epsf]{revtex}
\begin{document}
\draft
\title{A Mechanism for Lubrication between Surfaces with Atomic Level Roughness}
\author{J. B. Sokoloff}
\address{Physics Department and Center for Interdisciplinary Research on
Complex Systems, Northeastern University, Boston, MA 02115.}
\date{\today }
\maketitle

\begin{abstract}

It is proposed that lubricant molecules adsorbed on an interface between 
two asperities in contact, which is rough on the atomic scale,  can 
switch the interface from the strong to weak pinning regime, resulting in 
a large reduction in the static friction. This is proposed as a possible 
mechanism for boundary lubrication.
\end{abstract}

\begin{multicols}{2}
\narrowtext

In recent years there have been extensive studies, both in the form 
of experiment[1] and simulations[2],  of the behavior of 
liquids at an interface between two solid surfaces as the surfaces 
are squeezed together, which leads to the liquid molecules being partially 
driven out from the interface. In the high pressure limit, the 
liquid is reduced to a bilayer. One goal of these studies is to 
understand how boundary lubrication might occur. In boundary lubrication, 
there is a very thin layer of a lubricant which is able to both protect 
the surfaces against wear and to reduce the static and sliding friction. 
These studies, however, have not provided a physical mechanism for how 
such a thin layer of a liquid separating two surfaces is able to reduce 
friction. In this letter, I will propose a possible mechanism for the 
reduction of static friction by a thin liquid lubricant pressed between 
two disordered solid surfaces, based on the theory of collective 
pinning[3]. In collective pinning theory, 
there are two regimes There is a strong pinning regime, in which the elastic 
medium is able to distort a good deal so as to be able to nearly minimize 
the disordered potential, and there is a weak pinning regime, in which 
the elastic medium is only able to distort over a length L, known as the 
Larkin length, and as a consequence the interaction of the disordered 
potential with the elastic medium is only minimized within domains of 
length L, known as Larkin domains. Although 
we are interested in a disordered interface between two three 
dimensional solids, it is reasonable to expect that the 
study of an elastic solid in contact with a rigid disordered substrate 
will give results not qualitatively different from the problem of two 
elastic solids in contact at a disordered interface. 

Collective pinning theory has been applied to the friction problem by 
Caroli and Noziere[4] and by Persson and Tosatti[5] using a model for 
asperity interaction which puts the interface in the weak pinning limit
in which the Larkin length is very large. It was shown by the 
present author[6] that  
this model is at its critical dimension, which means that the interaction 
energy between the elastic solid and the substrate is minimized for 
a Larkin length comparable to 
the size of the interface for weak pinning and a Larkin length 
comparable to an atomic spacing in the strong pinning limit. 
For the case of a disordered interface 
between two solids, it is clear that the strength of the random 
potential due to the interface will increase as the two surfaces are 
pushed together with larger and larger forces, because the hard cores of 
the surface atoms are pushed together. When the potential is 
sufficiently strong, the system must switch over to the strong pinning 
regime, at which point, the Larkin length switches over from being 
comparable to the interface size to atomic dimensions[6]. In the 
weak pinning regime it was shown that the force of static friction 
per unit area decreases as the square root of the interface area, 
whereas in the strong pinning limit, it will be independent of the 
interface area. The mechanism proposed here for lubrication is that 
the lubricant molecules at the interface will get squeezed under pressure 
into regions of the surfaces which for un-lubricated surfaces were 
not in contact. By doing so 
the normal force pushing the two surfaces together gets supported 
over a larger area of contact. The reason that this can put 
the system into the weak pinning regime is that the elastic forces 
between neighboring pinning sites will be made larger relative to 
their interaction with the second surface.

Recently 
He, et. al., and Muser and Robbins[7] have shown for clean weakly interacting 
two dimensional 
incommensurate interfaces that the force of static friction per unit area 
falls to zero as 
$A^{-1/2}$ in the thermodynamic limit, where A is the interface area. 
Even identical solids are incommensurate if their crystalline axes are 
rotated with respect to each other. It is well known from studies of the 
one dimensional Frenkel-Kontorova model[1] that two incommensurate 
crystal lattices in contact with each other can slide with respect to 
each other with no static friction if their interaction is sufficiently 
weak compared to typical elastic energies, but if their mutual interaction 
is increased, above a critical value they make a transition (known as the 
Aubry transition [8]) to a state in which they do exhibit static friction. 
If two crystal lattices, consisting of atoms that do not interact chemically, 
are pushed together, their mutual interaction will 
increase because the hard core repulsions of the atoms in the two 
crystals will increase. Lancon[9] has studied the Aubry transition for 
a two dimensional model. Simulations by Muser and Robbins[7]  and by Muser[10] 
for two incommensurate surfaces pushed together with a normal force 
corresponding to pressures comparable to those occurring at contacting 
asperities at a typical sliding interface (i.e., of the order of a 10 GPa's) 
do not undergo an Aubry transition, and hence do not exhibit static friction.  
As discussed earlier, disorder, however, can pin contacting solids, 
just as it pins sliding 
charge density waves and vortices in a superconductor[3]. 

As a first step towards understanding how collective pinning theory, as described 
above, can provide a possible mechanism for boundary lubrication, let us 
consider the following likely but simple model for surface roughness at the 
sub-asperity level: Consider two 
identical three dimensional solids with atomically flat surfaces 
rotated with respect to each other, like those studied in Ref. 2. Let us 
now select atoms at random in the top layer of each surface and remove them. 
The atomic positions 
for this model of the top sub-layer of atoms are illustrated for a submonolayer
with half the number of atoms as occurs in a full monolayer in Fig. 1.
Now let us imagine placing two of these surfaces in 
contact and pressing them together with a load per unit area of the interface 
P. If the load 
is sufficiently high, we may safely assume that the force between a pair of 
surface atoms from the two surfaces which are in contact is dominated by the 
hard core interaction. Then, a good estimate of the magnitude of the mean force 
acting between a pair of surface atoms from the two surfaces that are in 
contact is P/n, where n is the number density 
per unit area of surface atoms. This is an estimate of the component of 
force normal to the interface, but since such atom pairs are rarely lined 
up so that one is exactly on top of the other, there will be components of 
force of comparable magnitude along the interface as well. 
Let us now imagine adding atoms to each of the outer layers until 
they each form a complete monolayer. Our model is then 
identical with the model studied in Refs. 6, 7, and 10. In this model it 
was reported in these references that even for loads per unit area as high 
as a GPa, there was no static friction between the surfaces. As the 
concentration of surface atoms is reduced from a complete monolayer on each 
surface by removing atoms chosen at random, there will certainly be a concentration 
at which the interface switches over from the weak pinning limit, in which 
the elastic energy dominates over the substrate potential (which 
in our case represents the second surface), to the strong pinning regime, 
in which the opposite is true. 

\begin{figure}
\centerline{
\vbox{ \hbox{\epsfxsize=7.0cm \epsfbox{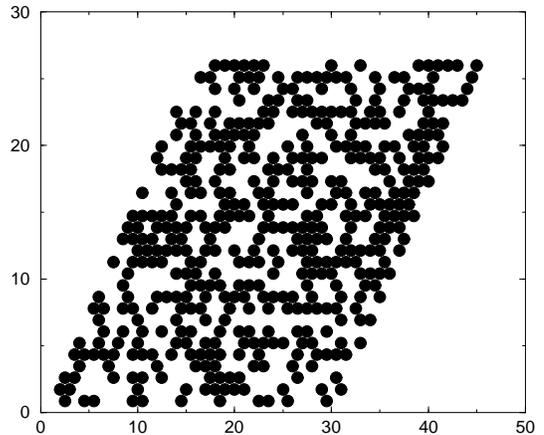}}
       \vspace*{1.0cm}
        }
}
\caption{The atomic positions are shown for a random submonolayer for a 
triangular lattice containing 
half of the number of atoms in a complete monolayer. The atomic positions are 
chosen at random. Distances are given in units of a lattice constant.}
\label{Fig1}
\end{figure}

Following Ref. 6, scaling methods, like those used by Fisher 
for charge density waves (CDW)[3], are used to study static friction for 
disordered interfaces. 
This can be accomplished by minimizing the potential energy of the solid in 
contact with a rigid disordered substrate at z=0 with 
respect to the size of a Larkin domain[3], which is expected to give 
qualitatively correct results for the problem of two elastic disordered 
solids in contact. Given that the energy density 
of the elastic solid is given approximately by
$$(1/2)E'|\nabla_t {\bf u}|^2+(1/2)E"|{\partial {\bf u}\over\partial z}|^2
+V({\bf r})\delta(z), \eqno (1)$$
where $E'$ and E" are elastic modulai,  
$V({\bf r})$ is the potential per unit area of the disordered substrate,  
${\bf u}({\bf r})$ is the local displacement of the solid, and 
$\nabla_t=({\partial\over\partial x},{\partial\over\partial y})$. Let us 
assume that $\nabla {\bf u}({\bf r})$ only differs significantly from zero 
over domains of length and width L parallel to the surface and height 
perpendicular to the surface equal to L'.  
Then, following the discussion in Ref. 6, the energy of a single domain 
is given by
$$E=(1/2)L'L^2 E'[|\nabla_t' {\bf u'}|^2/L^2+$$
$$|\partial {\bf u'}/\partial z'|^2/L'^2]
-V_0  c^{1/2}(L/a), \eqno (2)$$
where a is an atomic length scale (e.g., a lattice constant), c is the 
concentration of surface atoms in contact with the substrate and $V_0$ 
is a typical value of the substrate potential energy felt by an atom 
in contact with the substrate. In arriving at Eq. (2), we have assumed 
that since by definition of a Larkin domain, ${\bf u} ({\bf r})$ 
within a domain is small 
compared to a, the interaction of the substrate potential with a domain is 
random and hence the integral of this potential over a domain is proportional 
to the square root of the number of surface atoms in contact with the 
substrate (which is of the order of $c(L/a)^2$).  
Here, we assume that ${\bf u}(x,y,z)$ is 
equal to the function {\bf u'}(x',y',z'), where the function u' varies by an amount 
of the order of atomic length scales when x', y' and 
z', defined by (x',y',z')=(x/L,y/L,z/L'), each vary by an amount of order unity. 
The total potential energy due to the interaction of the solid with 
the substrate and the elastic distortion energy is the product of $A/L^2$, 
the number of domains along the interface, where A is the area of the 
interface, and Eq. (2). When the resulting expression is minimized 
with respect to L' one finds that $L'\approx L$ and the energy per unit 
area at the interface is given by
$$E/A\approx [(1/2)E' |\nabla'_t {\bf u'}|^2+$$
$$(1/2)E" |\partial {\bf u'}/\partial z'|^2
-V_0 c^{1/2}/a]/L, \eqno (3)$$
(where we use the average value of $|\nabla' {\bf u'}|^2$ here). By 
the above arguments concerning the hard core interactions, it is clear that 
the magnitude of the mean force exerted on a surface atom by the substrate $f_0$,  
is given by $nf_0\approx  P$. Then $V_0\approx Pa/n\approx Pa^3/c$, 
where a is a lattice constant. Then, the term in square brackets on the right 
hand side of Eq. (3) becomes $[(1/2)E' |\nabla'_t {\bf u'}|^2+
(1/2)E" |\partial {\bf u'}/\partial z'|^2-Pa^3/c^{1/2}]$. When this 
quantity is positive, the interface energy is minimized for large L (i.e., 
comparable to the interface size) and when the square bracketed expression is 
negative, it is minimized for small values of L (i.e., comparable to atomic 
dimensions).  Thus, it is clear that as c decreases, 
the interface will switch from weak pinning (if it was already in the weak 
pinning regime) to strong pinning. In the latter regime, by the arguments given 
in the last paragraph, the surfaces will be pinned together (i.e., there will 
be static friction). Thus for surfaces with sufficient roughness so that they 
are only in contact at a sufficiently dilute concentration of random points of 
contact, we are in the strong pinning regime, and hence, there must be 
sizable static friction. 

Let us now place molecules of some a lubricant on these surfaces,  
which attach themselves strongly to  
the surfaces.  This is the important property that these lubricant molecules 
must have in order to be good lubricants[11]. 
Then, let us assume that the attractive force between 
a lubricant molecule and a surface atom is much greater than the attraction 
between two lubricant molecules. These can be either 
single atoms or chains. If they are chains, we consider how the individual 
monomers position themselves on the interface. 
Simulations done for two flat surfaces in contact[2] show that under 
GPa pressures such a lubricant will get squeezed out until at the highest 
pressures we are left with a bilayer. In our case, where the surface is 
not smooth, we expect there to be a bilayer coating the steps (i.e., on the 
places on the outer layer at which there are surface atoms).  The 
regions in which there are no top submonolayer atoms present will also 
get filled in with lubricant molecules. The reason for this is that 
as the surfaces are squeezed together, some of the lubricant molecules 
that are driven from the steps will be pushed into these regions. 
I propose that this is a possible mechanism for boundary 
lubrication. This mechanism depends 
on there being a strong attraction between the surface and lubricant 
molecules, which causes the lubricant molecules to become trapped in 
one atomic layer deep "valleys" in the outer surfaces of the solids. 
We can see from Fig. 1 that the second layer of lubricant molecules 
that we propose to be present inside the valleys can easily be trapped 
there by lubricant molecules which are adsorbed on the tops of islands 
of top surface atoms (since they are assumed to be strongly attached to 
them). I have performed a simple Montecarlo calculation to illustrate 
this. A lubricant, consisting of 245 spherically symmetric molecules 
interacting with a Lennard-Jones potential is placed between the two 
surfaces. The lubricant molecules interact with each of the surfaces 
with a Steele potential [12]. Since when a molecule is sufficiently 
strongly pressed  
into one of the surfaces (because of the high pressures) the corrugation 
term in the Steele potential, which is the part that depends on x and 
y, dominates over the one that 
only depends on z, the potential can become unstable. This defect of 
the Steele potential is compensated for by making the corrugation 
term level off before it exceeds the term dependent only on z. 
The energy parameter $\epsilon$ in 
the Lennard-Jones 
potential $-4\epsilon [(\sigma/r)^6-(\sigma/r)^{12}]$ was chosen
to be a tenth of the energy parameter $\epsilon_{gs}$ 
in the Seele potential [12]. The potential minima are taken to lie on 
a triangular lattice of lattice spacing a=2.88 $A^o$. The bottom 
surface contains a hexagonally shaped hole of semi major and semi 
minor axes 7.6 and 6.6 $A^o$ respectively. The surfaces are circular with 
radii equal to $6a\sqrt(3)/2$ where a is taken to be 2.885 $A^0$. 
If $V_s (x,y,z)$ represents 
the potential due to this surface for values of x and y outside the hole 
(where the z-axis is normal to the interface), $V_s (x,y,z+z_0)$ is 
taken to be the potential inside the hole. Here $z_0$, the depth of 
the hole, is taken to be equal $a(2/3)^{1/2}$, the depth that the hole,  
resulting from removing atoms in the surface layer, would have on the 
surface of a hexagonal close packed (hcp) lattice with its c-axis 
normal to the 
surface or a face centered cubic lattice with its (111) surface parallel 
to the surface. The calculation is started with the lubricant molecules 
placed in an hcp lattice five atomic layers thick with lattice 
constant a between the two surfaces 
and centered over the hole. The initial separation of the surfaces is 
$15A^o$ which is just enough for the initial crystal of lubricant 
molecules to fit without being compressed. The surface is kept at 
that separation for $2\times 10^6$ iterations. The separation is then 
reduced by $0.5\times 10^{-7}$ for every Montecarlo iteration 
until the separation reaches 
$9.84A^o$.  The interface pressure at this separation was found to be  
$3.27\times 10^{11} dyn/cm^2$, which is larger than the pressure at the 
area of contact of two asperities used in Ref. 6. For smaller separations 
the total potential energy of the lubricant becomes large and positive, 
indicating that the film is becoming highly compressed. 
Results of these calculations for a $k_B T=20\epsilon$ 
(where $k_B$ is Boltzmann's constant and T is the absolute temperature) 
are shown in Fig. 2. The radius of the shaded 
sphere used to represent a lubricant molecule was chosen to be 
approximately equal to the radius used in the simulation in b and c 
but not in a (for clarity). As can be seen, the film gets compressed into 
a bi-layer outside of the hole and the hole gets filled  with a 
high concentration of lubricant molecules one monolayer thick, 
which could support load over the region in which the hole occurs. 
The interface between the two layers of lubricant will be an 
interface which will shear quite easily compared to an interface 
between unlubricated surfaces, as the only interaction 
acting across this interface is the interaction between pairs of lubricant 
molecules, which was chosen to be much weaker than the interaction 
between the two bare surfaces.
\begin{figure}
\centerline{
\vbox{ \hbox{\epsfxsize=4.0cm \epsfbox{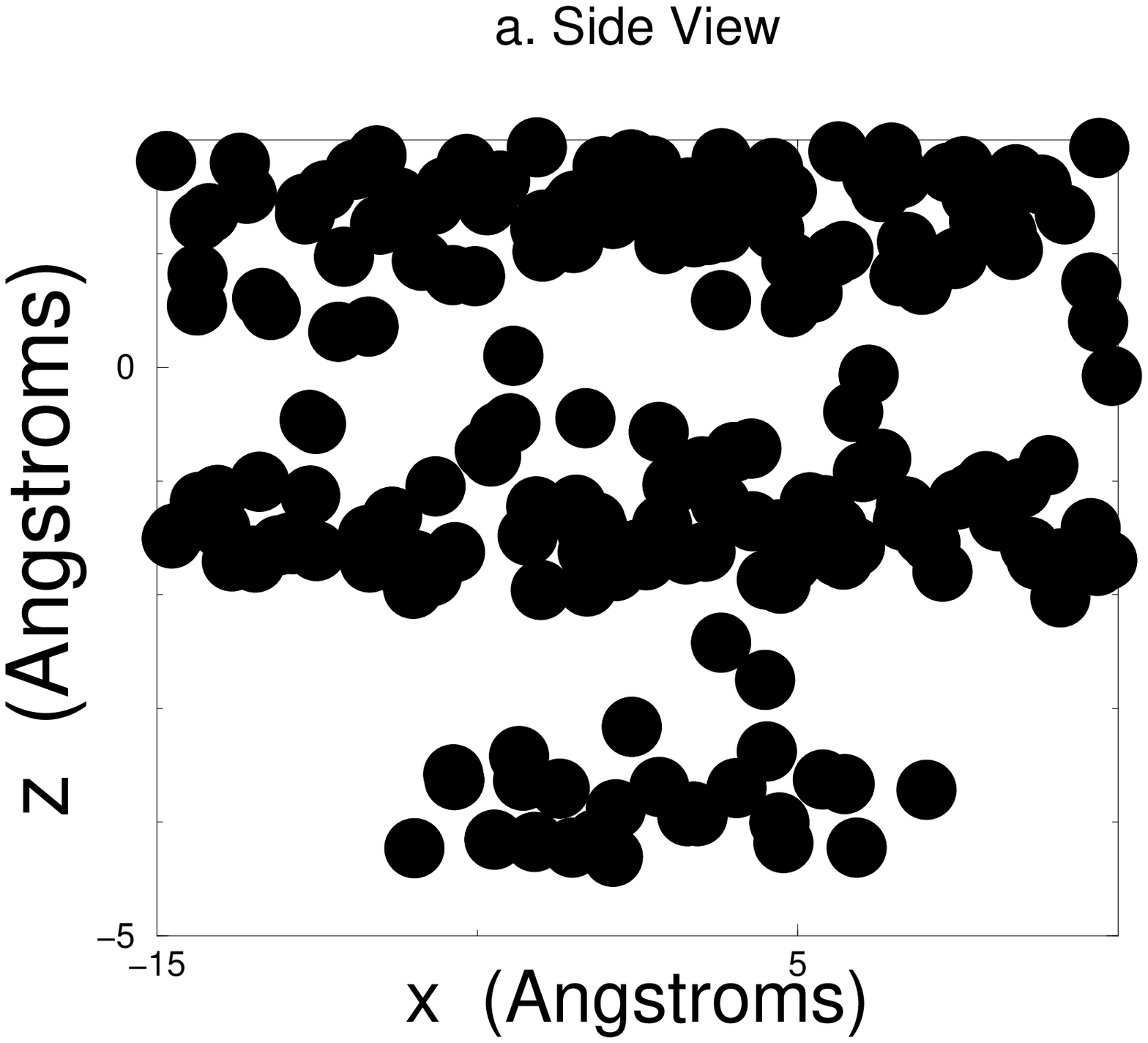}
                \hspace*{0.5cm} \epsfxsize= 4.0cm  \epsfbox{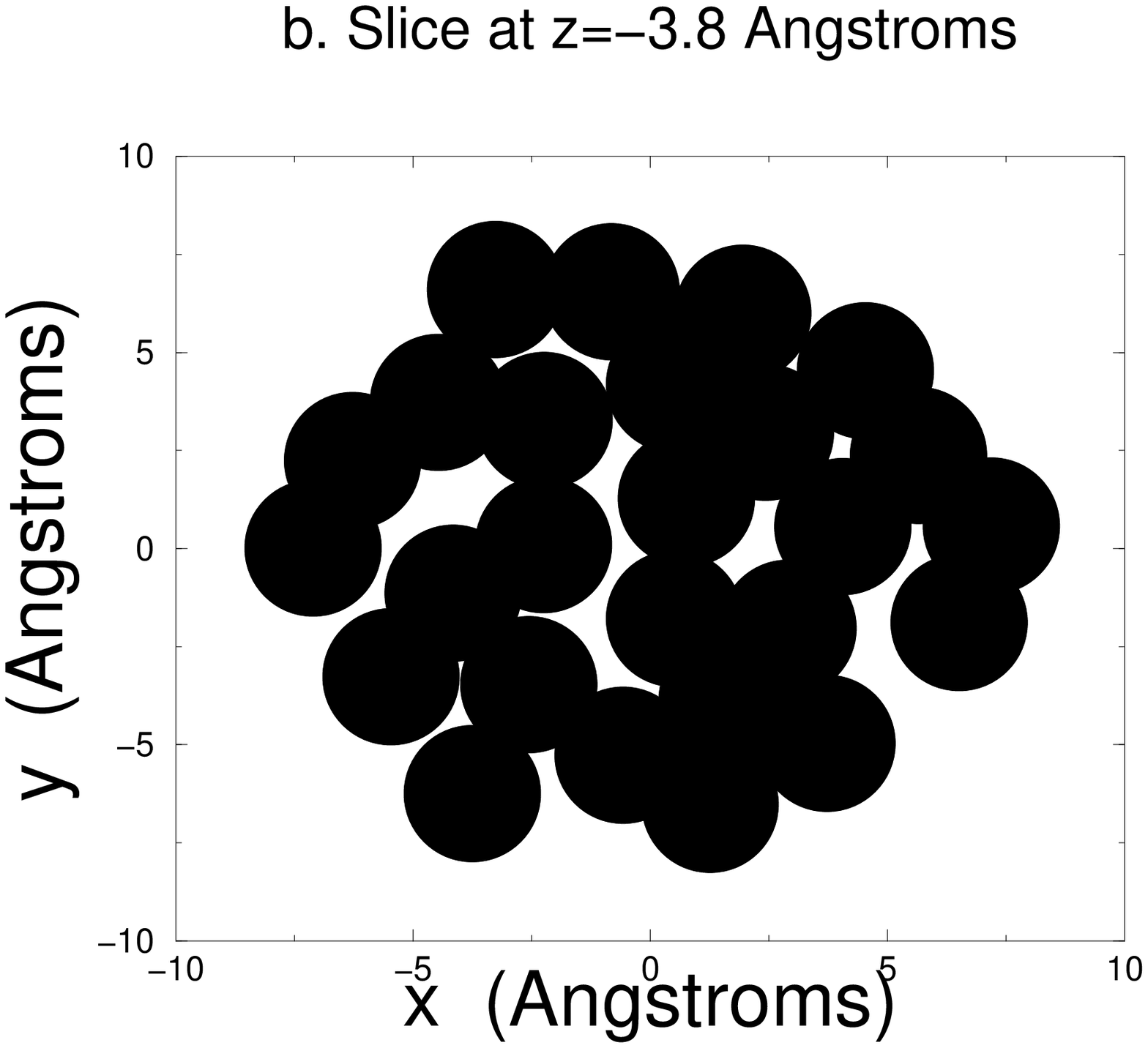} }
       \vspace*{1.0cm}
        } 
}

\vbox{ \hbox{\epsfxsize=4.5cm \epsfbox{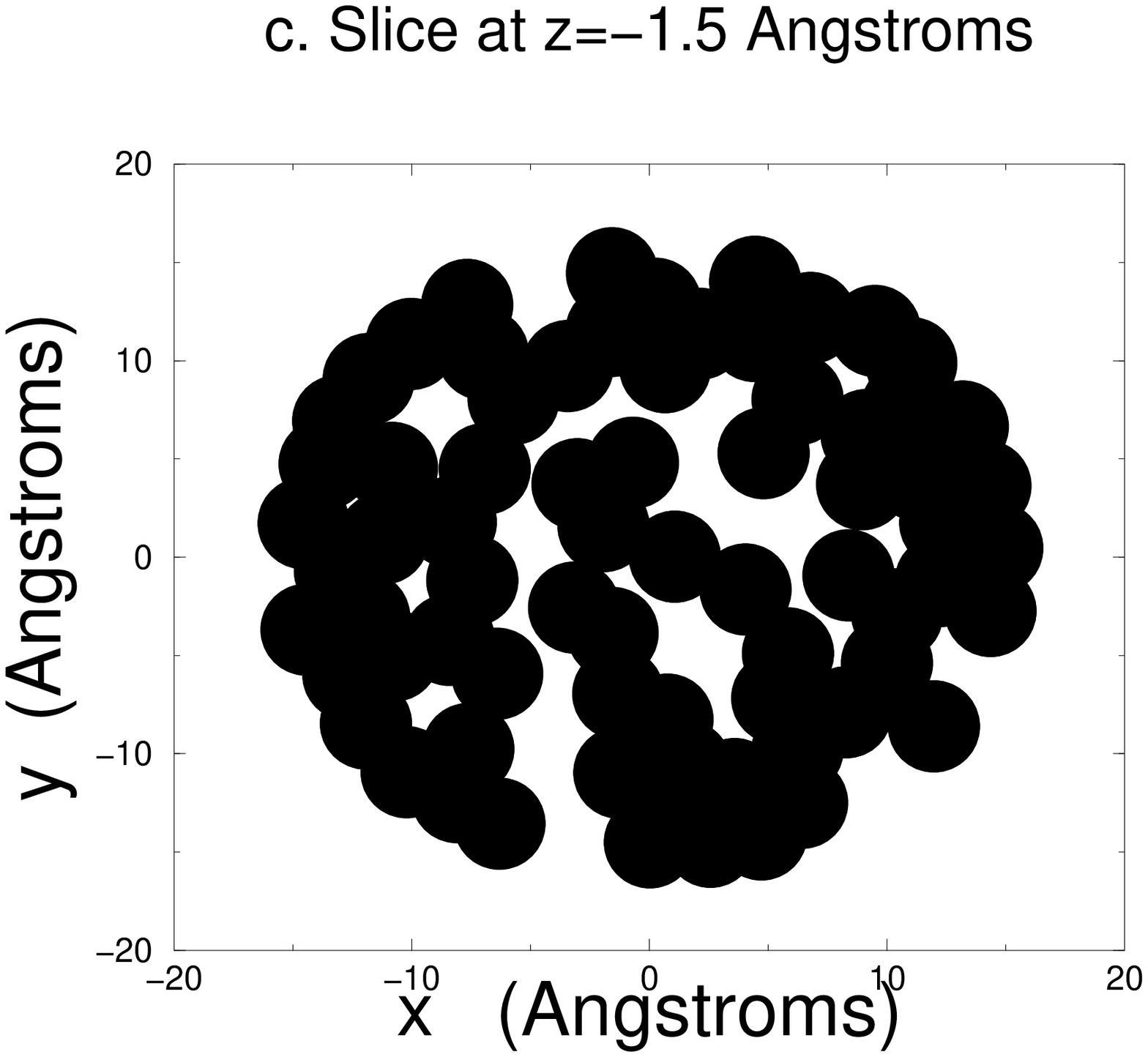}}
       \vspace*{1.0cm}
        }

\caption{In a, there is a side view of the distribution of molecules 
between the two surfaces. In b, the molecules in a slice of width equal 
to $1A^o$ centered around $z=-3.8A^o$ is shown, which are clearly located inside 
the hole. In c, the molecules in a slice of width equal 
to $1A^o$ centered around $z=-1.5A^o$ is shown, which are clearly located 
in one of the layers outside of the hole.}

\label{2}
\end{figure}

In contrast to the valleys in the outer surface assumed to occur at the 
interface between two asperities, lubricant molecules are not expected to get trapped in 
the space between micron scale asperities, because the depth of such 
regions is much too great to allow the attractive force between 
surface atoms and lubricant molecules to reach the lubricant molecules 
in most of this region. As a result, most of the lubricant found here 
remains liquid[2]. The mechanism for boundary lubrication suggested 
here should be applicable at the pressures that occur at the 
contact area between two asperities, which can reduce the lubricant 
concentration down to a monolayer or less[2].

The idea of lubricant molecules filling in valleys in the top layer of each 
surface and thus making the top surface more smooth at first sight seems 
like a familiar idea, but here we have provided a mechanism for how 
"smoothness" results in low friction. 
The idea that by doing so, the interface 
switches from the strong pinning regime, in which there is large static 
friction to the weak pinning regime in which there is little static 
friction[6,7] provides a mechanism for how such "smoothing" of the 
surface with lubricant molecules can lead to low friction.
\acknowledgments

I wish to thank the Department of Energy (Grant DE-FG02-96ER45585).

\end{multicols}{2}

\end{document}